\documentclass[twocolumn,notitlepage,floats,nofootinbib,amsmath,amssymb,aps,prd]{revtex4-1}

\usepackage{amsmath,amssymb,amsfonts,latexsym,cancel}

\usepackage{mathtools}
\usepackage{graphicx}
\usepackage{color}
\usepackage{amsbsy}
\usepackage{hyperref}
\usepackage{tikz}

\usepackage{seqsplit}

\usepackage{comment}

\newcommand{\be}{\begin{equation}}
\newcommand{\beq}{\begin{equation}}
\newcommand{\ee}{\end{equation}}

\def\bea {\begin{eqnarray}}
\def\eea {\end{eqnarray}}

\def\f{\frac}
\def\tf{\tfrac}
\def\nn{\nonumber}
\def\lp{\ell_{\rm Pl}}

\def\dd{{\rm d}}

\usepackage{amsthm}

\theoremstyle{plain}

\newtheorem*{theorem*}{Theorem}

\definecolor{newgreen}{rgb}{0.0, 0.75, 0.0}
\definecolor{cadmiumgreen}{rgb}{0.0, 0.42, 0.24}

\begin{document}

\title{Shell-crossings and shock formation during gravitational collapse \\ in effective loop quantum gravity}

\author{Francesco Fazzini} \email{francesco.fazzini@unb.ca}
\affiliation{Department of Mathematics and Statistics, University of New Brunswick, \\
Fredericton, NB, Canada E3B 5A3}

\author{Viqar Husain} \email{vhusain@unb.ca}
\affiliation{Department of Mathematics and Statistics, University of New Brunswick, \\
Fredericton, NB, Canada E3B 5A3}

\author{Edward Wilson-Ewing} \email{edward.wilson-ewing@unb.ca}
\affiliation{Department of Mathematics and Statistics, University of New Brunswick, \\
Fredericton, NB, Canada E3B 5A3}

\begin{abstract}

Effective models of gravitational collapse in loop quantum gravity for the Lema\^itre-Tolman-Bondi spacetime predict that collapsing matter reaches a maximum finite density, bounces, and then expands outwards. We show that in the marginally bound case, shell-crossing singularities commonly occur for inhomogeneous initial profiles of the dust energy density; this is the case in particular for all profiles that are continuous and of compact support, including configurations arbitrarily close to the Oppenheimer-Snyder model. When a shell-crossing singularity occurs, it is necessary to seek weak solutions to the dynamics; we argue that weak solutions typically contain shock waves.

\end{abstract}

\maketitle

\section{Introduction}
\label{s.intro}

Recent work on the spherically symmetric gravity-dust Lema\^itre-Tolman-Bondi (LTB) spacetimes \cite{Bojowald:2008ja, Bojowald:2009ih, Tibrewala:2012xb, Kelly:2020lec, Alonso-Bardaji:2020rxb, Giesel:2021dug, Husain:2021ojz, Husain:2022gwp, Giesel:2022rxi, Giesel:2023tsj} in effective loop quantum gravity (LQG) has provided a class of models with a local degree of freedom for studying quantum gravity effects in the gravitational collapse to a black hole \cite{Husain:2021ojz, Husain:2022gwp, Giesel:2022rxi, Giesel:2023hys}.  (For recent reviews on black holes in LQG, see \cite{Gambini:2022hxr, Ashtekar:2023cod}.)

These studies agree that quantum gravity corrections only become important when the energy density of the collapsing dust field nears the Planck scale, at which point LQG effects lead to a repulsion that causes the initially in-falling dust field to undergo a non-singular bounce. However, a major difference is that some models predict the formation of a shock (during or before the bounce) \cite{Kelly:2020lec, Husain:2021ojz, Husain:2022gwp}, while others find that no shocks are formed \cite{Fazzini:2023scu, Giesel:2023hys}. Our purpose is to analyze this difference. Before doing so it is useful to first review the situations, in general, where shock wave formation becomes possible.

The LTB model is a system with one local degree of freedom. Its canonical equations of motion are a pair of coupled first-order non-linear partial differential equations; this is the case for both classical general relativity and the effective LQG models. It is therefore a useful non-perturbative model for gravitational collapse as its dynamics remain valid even in regions of large inhomogeneity.

A powerful tool to solve such equations is the method of characteristics \cite{PDEbook}. This method can be demonstrated by considering the equation
\beq \label{pde}
\partial_t u(x,t) + v(x,t;u) \partial_x u(x,t) = 0;
\ee
this is a non-linear advection equation where the velocity $v$ of the field $u$ depends on space, time and the field itself.

The method of characteristics defines curves $x(s)$ and $t(s)$, (called characteristics) in the $x-t$ plane that determine the evolution of $u$ along the curves through $\dd u / \dd s$. For an advection equation of the form \eqref{pde}, this gives
\beq
\f{\dd u}{\dd s} =
\partial_t u \cdot \f{\dd t}{\dd s} + \partial_x u \cdot \f{\dd x}{\dd s},
\ee
and reproduces \eqref{pde} provided  
\beq
\label{char-pde}
\f{\dd t}{\dd s} = 1, \qquad \f{\dd x}{\dd s} = v(x,t,u);
\ee
it follows that $u$ is constant along such curves,
\beq
\f{\dd u}{\dd s} = 0.
\ee
The problem is then reduced to solving the ordinary differential equations for $x(s)$, one for each $x$, with the initial conditions $x(s_0)=x_0$ and $t(s_0)=t_0$; the two other sets of equations for $\dd u / \dd s$ and $\dd t / \dd s$ are easily solved: $u(s)=u(x_0, t_0) = u_0$ and $t=s - s_0 + t_0$.

This method  can be viewed as a change of coordinates from $(x,t)$ to $(X, s)$, where $X$ labels the characteristic curves that are each parametrized by $s$. (Typically, $X$ is taken to be $X = x_0$; although convenient this choice is not essential.)  The new $X$ coordinate is chosen specifically so that it is comoving, in the sense that it follows the field $u$ such that $u$ remains constant along curves of constant $X$.

The main drawback of the method  is that the solutions hold only up to the points where the characteristic curves $x(t)$ intersect---since $v$ is field dependent, characteristic crossing is a possibility (unlike for constant $v$).  (If one attempts to use the method of characteristics beyond the point that characteristics cross, the resulting ``solution'' for $u$ is multi-valued, hence not a function.) It is possible to check whether characteristics cross by calculating the Jacobian of the coordinate transformation between $(x,t)$ and $(X,s)$: the Jacobian vanishes when characteristics cross. In particular, for the case that $s=t-t_0 + s_0$ considered here, the Jacobian vanishes if and only if $\partial_X x = 0$.

If characteristics cross, then the solution for $u(x,t)$ is not unique after the crossing point and in such cases it is necessary to look for weak solutions. In particular, in the weak solution, shocks typically form if characteristic curves intersect, while rarefaction waves arise if characteristic curves separate.

Weak solutions solve the integrated form of the equation of motion; for example, a weak solution $u_W$ for \eqref{pde} on the domain $D=\mathbb{R}\times[0,\infty)$ with  initial data $u(0,x)$ is one that satisfies
\be \label{integraleq}
\int_D \dd x \, \dd t \ u_W \left[\partial_t \phi + \partial_x(\phi v)\right]  =\int_\mathbb{R} \dd x \  \phi(x,0) u(0,x),
\ee
for all  smooth functions $\phi(x,t)$ of compact support on $D$, assuming $v = 0$ on the boundary $\partial D$ of the domain of interest. It is evident that weak solutions $u_W$ are not required to be differentiable on $D$, and therefore need not be solutions of \eqref{pde}. 

If comoving coordinates are utilized from the start for a system with dynamics given by a non-linear  equation, then the resulting equations of motion are ordinary differential equations identical to the parametric equations for the characteristic curves, and the defining non-linear wave equation does not appear explicitly. This apparent shortcut is useful as long as the characteristic curves do not cross; however, if they do cross, the implication is that the chosen comoving coordinates have failed. It then becomes necessary to recast the system in terms of the original non-linear PDE in another coordinate system and search for weak solutions.

In Sec.~\ref{s.ltb} we describe the dynamics for LTB spacetimes as PDEs and as characteristic equations; we discuss the conditions for shell-crossing singularities and their relation to characteristic crossing. We next consider LTB configurations that describe gravitational collapse, focusing on the Oppenheimer-Snyder (OS) model in Sec.~\ref{s.os}; in Sec.~\ref{s.beyond} we consider more general dust profiles, in particular we show that shell-crossing singularities commonly occur, including for all profiles that are initially marginally bound, continuous and of compact support. We conclude in Sec.~\ref{s.shock} with a discussion on the physics of shocks in gravitational collapse drawing on insights from fluid mechanics where shocks are ubiquitous.

\section{LTB characteristic equations}
\label{s.ltb}

In this section we show that the LTB equations in comoving coordinates are the characteristic equations of a pair of non-linear PDEs of the type~(\ref{pde}). It then follows that shell-crossing singularities corresponding to the crossing of characteristic curves occur, a result in agreement to that established in the areal gauge \cite{Husain:2021ojz, Husain:2022gwp}. These observations highlight the fact that shock waves arising in weak solutions are not gauge artifacts.

The LTB metric in the comoving coordinate $R$is
\beq \label{metric}
\dd s^2 = - \dd t^2 + \f{(\partial_R r)^2}{1 + \mathcal{E}} \dd R^2 + r^2 \dd\Omega^2,
\ee
where the areal radius $r=r(R,t)$ and spatial curvature $\mathcal{E}=\mathcal{E}(R,t)$ are to be determined by either the classical Einstein equations or a counterpart with quantum corrections.  

The gravitational mass $m$ up to radius $R$
\beq \label{def-m}
m(R, t) = 4 \pi \int_0^R \!\! \dd \tilde R \,\, r(\tilde R, t)^2 [\partial_{\tilde R} r(\tilde R, t) ] \, \rho(\tilde R, t),
\ee
also plays an important role in the LTB dynamics;  inverting this relation gives the formula for the dust energy density $\rho$,
\beq
\rho = \f{\partial_R m}{4 \pi r^2 \partial_R r}.
\label{rho}
\ee

With these definitions, the effective LQG dynamical equations in the comoving coordinates  are \cite{Giesel:2023hys}
\begin{gather}
\label{e-m-dot}
\f{\partial m(R,t)}{\partial t} = 0, \qquad
\f{\partial {\mathcal E}(R,t)}{\partial t} = 0, \\ 
\label{rdot}
\left( \f{\dot r}{r} \right)^2 = \left( \f{2 G m}{r^3} + \f{\mathcal{E}}{r^2} \right)  \left[ 1 - \Delta \left( \f{2 G m}{r^3} + \f{\mathcal{E}}{r^2} \right) \right], 
\end{gather}
where a dot denotes a derivative with respect to $t$, the Barbero-Immirzi parameter is set to unity, and $\Delta \sim \lp^2$ is the minimum eigenvalue of the area operator in LQG; the limit $\Delta\to0$ gives the classical equations.

The first two equations show that $m=m(R)$ and $\mathcal{E}= \mathcal{E}(R)$; these functions are fixed by the initial data.

The third equation \eqref{rdot} is an infinite set of ordinary differential equations, one for each value of the coordinate $R$---these correspond to the characteristic equations for the curves $r(t)$ that are comoving with the dust. These are to be solved with initial conditions $r(R,t_0)=h(R)$, where $h(R)$ specifies how the comoving coordinate $R$ is related to the areal radius $r$ at $t=t_0$. If the areal radius is monotonically increasing at $t=t_0$, it is convenient (though not required) to rescale the comoving coordinate $R$ such that $h(R) = R$, i.e., that $r(R,t_0) = R$.

A full solution of Eqs.~\eqref{e-m-dot}--\eqref{rdot} is therefore specified by three functions 
\be
r_R(t)= r(t, R; m(R),\mathcal{E}(R),h(R)).
\ee
Such a large class of solutions opens up the possibility that with appropriate  initial conditions,  solutions $r_k(t,R;R_k)$, where $k$ labels different characteristics, intersect in the $r-t$ plane. While it may be possible to avoid  characteristic crossing if the functions $m,\mathcal{E},h$ are carefully chosen, this will not be the case in general. (For an analysis of this for LTB spacetimes in classical general relativity, see \cite{Hellaby:1985zz}.) Indeed, we show in the following sections that the characteristics determined by \eqref{rdot} do cross for a wide range of initial data. 

The formula \eqref{rho} for the dust energy density indicates a divergence at the points where $\partial_R r=0$ and $\partial_R m \ne 0$, and it can be checked that curvature invariants also diverge at such points \cite{Hellaby:1985zz}; these are shell-crossing singularities. Shell-crossing singularities are weak singularities in the sense that neighbouring test particles remain separated \cite{Szekeres:1995gy} rather than being crushed together or ripped infinitely far apart as for a strong singularity. Since weak singularities are not cured in loop quantum cosmology \cite{Singh:2009mz}, it is not surprising that they may arise in effective LQG models of black hole collapse as well.

Importantly, the first condition $\partial_R r=0$ for a shell-crossing singularity also signals that characteristics have crossed. This is because the  Jacobian for the coordinate transformation to the comoving coordinate $R$ vanishes when $\partial_R r=0$ (in the same way that $\partial_X x = 0$ signals characteristics crossing in the example presented in Sec.~\ref{s.intro}). Therefore, a key point worth emphasizing is that for LTB spacetimes, a shell-crossing singularity signals that characteristics cross precisely at the location of the singularity. Thus when such a point is reached, it is necessary to allow for weak solutions, and the concomitant expectation of shock formation. (In contrast, divergence of characteristics gives a rarefaction wave.)

\begin{figure*} 
\begin{center}
   \includegraphics[width=\textwidth]{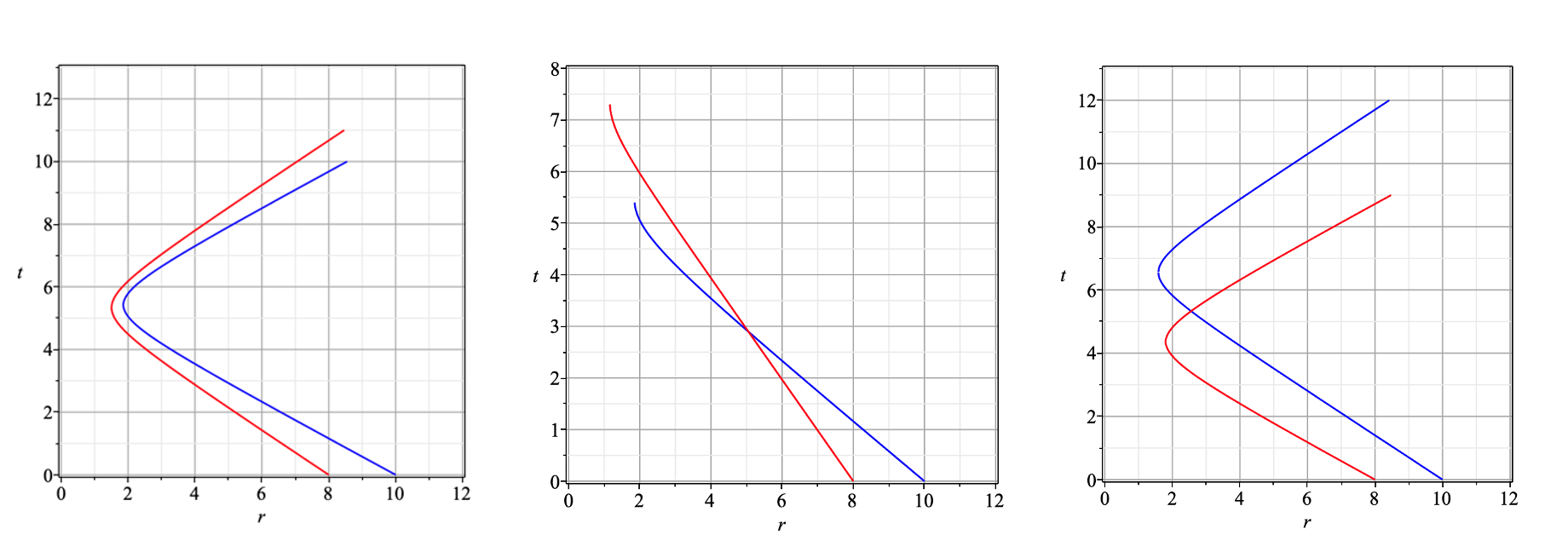}
  \caption{Examples of characteristic curves for infalling dust from solutions of \eqref{rdot}: from left to right, these show no crossing, crossing before the bounce, and crossing of a curve that has bounced with an ingoing one. At the point where characteristics cross, it becomes necessary to consider weak solutions. For $\partial_R m \neq 0$, shell-crossing singularities and  characteristic crossing coincide, showing that these are not coordinate artefacts. The initial conditions for the curves shown here (with $\Delta =1$) are: for the red curve, $r(0)=8,\ m(8)=0.2$ (for all three plots) and from left to right ${\mathcal E}(8)=2,1,3$; for the blue curve, $r(0)=10,\ m(10)=0.4$ (for all three plots), and from left to right ${\mathcal E}(10)=3,3,2$.}
  \label{fig1}
\end{center}
\end{figure*}

Examples of characteristic curves for the characteristic equation \eqref{rdot} are shown in Fig.~\ref{fig1} (with the initial conditions for each of the three plots given in the caption): the leftmost plot shows two such curves that do not intersect, the central plot shows two that intersect before the bounce, and the rightmost plot shows two that cross after the inner one has bounced.

Since the characteristics are curves of constant $R$, at an intersection, either $R$ or $r$ must fail as a coordinate: is this intersection in the $r-t$ plane a single spacetime point (in which case $R$ is not a good coordinate), or are they separate points $R_1$ and $R_2$ that have the same areal radius (in which case $r$ is not a good coordinate)?  The discussion above concerning the method of characteristics indicates that when characteristics cross, it is the comoving coordinate (in this case $R$) which fails. This is made particularly clear in the case of LTB spacetimes by the presence of a shell-crossing singularity if $\partial_R m \neq 0$: the divergence in $\rho$ is due to shells of dust crossing each other, i.e., being at the same spacetime point, as is correctly described by the areal radius coordinate $r$, but not by the comoving coordinate $R$--- therefore $R$ fails at these points. (For further discussion on the limitations of the method of characteristics for LTB spacetimes in classical general relativity, see \cite{Nolan:2003wp, Lasky:2006hq}.)

When characteristics cross, as in the examples shown in the middle and rightmost plots of Fig.~\ref{fig1}, it becomes necessary to look for weak solutions.
Weak solutions are derived from the integral form of the PDE from which the characteristic equations are derived. We thus seek the PDEs that give \eqref{e-m-dot}-\eqref{rdot} as their characteristic equations. This is readily accomplished by noting that the areal radius $r$ in \eqref{rdot} plays the role of $x$ in \eqref{char-pde}, and the fields are $m$ and $\mathcal{E}$. Changing variables from $(R, t)$ to $(r, t)$, and using the chain rule gives
\begin{gather}
\label{m-pde}
\partial_t m + f(r,m,\mathcal E) \, \partial_r m = 0, \\
\label{e-pde}
\partial_t \mathcal{E} + f(r,m,\mathcal E) \, \partial_r \mathcal{E} = 0,
\end{gather}
where the velocity of both fields comes from \eqref{rdot},
\beq \label{vel-ltb}
f^2 = r^2 
\left( \f{2 G m}{r^3} + \f{\mathcal{E}}{r^2} \right)  \left[ 1 - \Delta \left( \f{2 G m}{r^3} + \f{\mathcal{E}}{r^2} \right) \right] .
\ee
A direct calculation shows that the characteristic equations for these coupled partial differential equations are precisely \eqref{e-m-dot}--\eqref{rdot}; this is shown in App.~\ref{app.pg}.

Interestingly, this analysis also suggests that the most appropriate radial coordinate to use for LTB weak solutions is the areal radius, as was used in earlier work both in general relativity \cite{Lasky:2006hq} and effective LQG \cite{Husain:2021ojz, Husain:2022gwp}.

Since the equations above arise from an effective dynamics that is invariant under spatial diffeomorphisms \cite{Giesel:2023tsj}, it is possible to use other radial coordinates. However, since the equations hold specifically in the dust-time gauge, obtaining a quantization of this model where the quantum constraint algebra is fully realized remains an open question; for various approaches to the problem see, e.g., \cite{Bojowald:2009ih, Tibrewala:2012xb, Alonso-Bardaji:2020rxb, Han:2022rsx, Bojowald:2024beb}.

As a final comment, we note  that a missing ingredient in \eqref{vel-ltb} is the sign of $f$. This suggests that $m$ and $\mathcal{E}$ are not the best of choice of fundamental fields to consider in the LTB spacetime, since these do not determine ${\rm sgn}(f)$. Indeed, it turns out that using a component of the Ashtekar-Barbero connection (instead of $m$) solves this problem; details appear in App.~\ref{app.pg}.

\section{Oppenheimer-Snyder Collapse}
\label{s.os}

In this section we revisit the Oppenheimer-Snyder (OS) collapse model \cite{Oppenheimer:1939ue} as a special case of the LTB spacetime in the effective LQG framework, with remarks on the use of generalized Painlev\'e-Gullstrand and comoving coordinates, and on the dust time gauge.

The collapsing OS star is assumed to have  a radially constant energy density, vanishing pressure, with vacuum outside; the initial condition for $\rho$ is
\beq \label{rho-os}
\rho = \begin{cases}
\rho_o, \qquad & {\rm for~} R < R_{OS}, \\
0, \qquad & {\rm for~} R > R_{OS},
\end{cases}
\ee
where $R_{OS}$ is the location of the surface of the OS `star' in terms of the comoving radial coordinate $R$.  

Due to the fact that the interior is (a portion of) the homogeneous and isotropic cosmological Freidman-Lema\^itre-Robertson-Walker (FLRW) spacetime, it is possible to study LQG effects by assuming the interior dynamics are those given by loop quantum cosmology for a dust-filled FLRW spacetime, and use matching conditions to determine the exterior \cite{Lewandowski:2022zce, Bobula:2023kbo, Fazzini:2023scu} (see also \cite{Hossenfelder:2009fc, BenAchour:2020gon} for earlier work on OS collapse in LQG). Alternatively, the OS model can also be seen as a solution to the LTB spacetime and derived from the equations of motion given above \cite{Kelly:2020lec, Husain:2022gwp, Giesel:2022rxi, Giesel:2023hys}.

The exterior vacuum solution is the same in all cases (and matches what was found in studies of effective LQG in vacuum spherical symmetry \cite{Gambini:2020nsf, Kelly:2020uwj}), as are the dynamics of the interior, but the matching between the interior and exterior differs subtly in three approaches described in the following; this is because the matching has to be done along a timelike 3-surface of topology $S^2 \times \mathbb{R}$, where the two sphere must have a certain areal radius as a function of time, and there are multiple possible 3-surfaces in (the maximal extension of) the vacuum exterior for which this matching is possible.

\subsection{Generalized Painlev\'e-Gullstrand coordinates}

The generalized Painlev\'e-Gullstrand coordinates for LTB spacetimes are obtained by using the areal radius $r$ as the radial coordinate (this requires that, at all times, the areal radius increases monotonically from the origin to infinity), and using the dust field as a relational time variable; these are known as the areal and dust-time gauges respectively. In these coordinates, the metric is
\beq
\dd s^2 = -\dd t^2 + \f{1}{1 + \mathcal{E}} (\dd r + N^r \dd t)^2 + r^2 \dd \Omega^2,
\ee
and the equations of motion in these coordinates are derived in \cite{Kelly:2020lec, Husain:2021ojz, Husain:2022gwp} and are summarized in App.~\ref{app.pg}. Considering the marginally bound case $\mathcal{E} = 0$, the solution for the OS configuration is \cite{Kelly:2020lec, Husain:2022gwp}
\beq
N^r = \begin{cases}
\displaystyle \f{-6 rt}{9t^2 + 4 \Delta}, \quad & {\rm for~} r < L(t), \\[8pt]
\displaystyle \sqrt{ \f{R_S}{r} \left( 1 - \f{\Delta R_S}{r^3} \right) }, \quad & {\rm for~} r > L(t),
\end{cases}
\ee
where $R_S = 2GM$, with $M$ the total gravitational mass of the OS star, and
\beq
L(t) = \left( \f{9 R_S t^2}{4} + \Delta R_S \right)^{1/3}
\ee
is the areal radius of the OS star.

Note that the shift is negative for $t < 0$, both in the interior and exterior regions, and it is straightforward to check that $N^r$ is in fact continuous for $t \le 0$. On the other hand, in the interior region (but not in the exterior region) the shift changes sign after the bounce at $t=0$, so a discontinuity forms in $N^r$ for $t>0$. This discontinuity is captured in a component of the Ashtekar-Barbero connection as well, which also becomes discontinuous after the bounce. Further, it can be verified that characteristics cross at the bounce, indicating the necessity of looking for weak solutions.

A detailed study of weak solutions, for the OS model as well as other configurations, shows that a shock forms at the latest at the bounce \cite{Husain:2021ojz, Husain:2022gwp}. This analysis (including the result concerning the formation of the shock), however, depends on the areal and dust-time gauges being valid. In the next subsections, we review how the results (specifically for the OS  model) may differ if one or the other of these gauges is relaxed.

\subsection{Relaxing the areal gauge}
\label{s.os-ch}

It is possible to avoid the areal gauge by instead using the so-called LTB gauge \cite{Bojowald:2008ja, Giesel:2023tsj}, which selects coordinates that are comoving with the energy density of the dust, as reviewed in Sec.~\ref{s.ltb}, giving the diagonal metric \eqref{metric}. 

For the OS initial conditions \eqref{rho-os},
\beq
m = \begin{cases}
\displaystyle \f{4 \pi G \rho_o R^3}{3}, \qquad &{\rm for~} R < R_{OS}, \\[8pt]
\displaystyle \f{4 \pi G \rho_o R_{OS}^3}{3}, \qquad &{\rm for~} R > R_{OS}.
\end{cases}
\ee
using the initial condition $r(R, t_0) = R$. Although this second condition is not strictly necessary, it simplifies the calculations (in particular, when imposing the initial conditions it is not necessary to worry about the $\partial_R r$ term in the relation between $\rho$ and $m$, although the $\partial_R r$ term cannot be neglected for $t \neq t_0$).

Due to the use of comoving coordinates, the dynamics are given by the ordinary differential equation \eqref{rdot} for the characteristic curves $r(R, t)$, that can be solved analytically with the result (for $\mathcal{E} = 0$) that \cite{Giesel:2023hys}
\beq
r(R,t) = \left( 2Gm \left( \f{9}{4} (t - \alpha)^2 + \Delta \right) \right)^{1/3},
\ee
where there is a constant of integration $\alpha$ for each $R$.

The initial condition $r(R, t_0) = R$ also determines the constants of integration $\alpha$. For the interior $R < R_{OS}$,
\beq
\alpha(R) = t_0 + \sqrt{ \f{4}{9} \left( \f{3}{8 \pi G \rho_o} - \Delta \right) },
\ee
note that these $\alpha$ are independent of $R$, while for the exterior $R > R_{OS}$,
\beq
\alpha(R) = t_0 + \sqrt{ \f{4}{9} \left( \f{R^3}{2 G M} - \Delta \right) },
\ee
where $M = m(R_{OS})$ is the total gravitational mass.

Given this solution for the OS model, it is possible to explicitly check whether a shell-crossing singularity forms or not by finding all occasions where $\partial_R r = 0$, and checking to see if $\partial_R m \neq 0$ at that particular value of $R$.

For the interior, recalling that $\alpha$ is independent of $R$,
\beq
\partial_R r = \left( \f{8 \pi G \rho_o}{3} \left( \f{9}{4} (t - \alpha)^2 + \Delta \right) \right)^{1/3},
\ee
and it is clear that $\partial_R r$ is always non-zero.

For the exterior, it is $m$ that is independent of $R$, and
\beq
\partial_R r = \f{3 G M}{r^2} \, (t - \alpha) \, \partial_R \alpha.
\ee
The prefactor as well as $\partial_R \alpha$ are always finite so, to conclude, $\partial_R r = 0$ only in the exterior, and at exactly one instant of time for each $R \ge R_{OS}$, specifically
\beq
t = t_0 + \sqrt{ \f{4}{9} \left( \f{R^3}{2 G M} - \Delta \right) }.
\ee
Despite this, it is clear that there is no shell-crossing singularity, since $\partial_R m = 0$ in the exterior region and therefore $\rho=0$ in the exterior region at all times.

Nonetheless, this calculation might give us pause: in a sense, a shell-crossing singularity is avoided because the energy density drops suddenly to 0 at the boundary $R=R_{OS}$. Note that for the exterior, $\partial_R r = 0$ also at the boundary $R=R_{OS}$, occurring exactly at the time when the interior bounces (which is the time when a shock is found to form according to the wave equation expressed in generalized Painlev\'e-Gullstrand coordinates)---this may feel a little too close for comfort.  Indeed, it is natural to ask what happens if instead $\rho$ decreases continuously to zero at the boundary of the star. Could it be possible for $\partial_R r$ to vanish in the region where $\rho$ is decreasing (continuously, although perhaps rapidly)? As shall be shown in Sec.~\ref{s.beyond}, the answer is yes: even for configurations that are arbitrarily close to Oppenheimer-Snyder, if $\rho$ is continuous and of compact support then a shell-crossing singularity necessarily occurs, at the latest at a time $\sim \sqrt\Delta$ after the bounce, at which point the comoving coordinates \eqref{metric} fail and it is necessary to find weak solutions to the dynamics. 

\subsection{Relaxing the dust-time gauge}

An alternative possibility is to instead relax the dust-time gauge, by allowing the dust field (used as a relational clock here) to evolve at different rates in the interior and exterior regions, as considered in \cite{Fazzini:2023scu}. With this change, it is possible to impose the Israel junction conditions to obtain a continuous geometry, in the sense that the induced metric on the boundary, and the extrinsic curvature on the boundary, as calculated from the interior and the exterior give the same result, and since the geometry resulting from this construction is continuous there is no shock in this case.

The underlying idea in this process is to change the (relational) time coordinates after the bounce in order to avoid the crossing of characteristics, and hence the necessity of looking for a weak solution. The result is an Oppenheimer-Snyder collapse model without the formation of any shocks in the geometry or in the dust energy density, although at the expense of a discontinuity in the dust field (the relational time variable) itself.

\section{Beyond Oppenheimer-Snyder}
\label{s.beyond}

In this section, we consider general initial conditions where the initial profile in $\rho$ is continuous (unlike OS where there is a jump discontinuity in the energy density at $R=R_{OS}$).  We avoid using the areal gauge, instead using the metric \eqref{metric} and the equation of motion \eqref{rdot} for $r(R,t)$ to describe the dynamics.  The resulting dynamics, as is shown in detail below, lead to the formation of shell-crossing singularities even for collapse models whose initial profile for $\rho$ is arbitrarily close to the OS profile. 

\subsection{Conditions for shell-crossing singularities}
\label{s.cond}

Given a specific mass function $m(R)$ and spatial curvature $\mathcal{E}(R)$, it is possible to calculate whether the dynamics will lead to the occurrence of a shell-crossing singularity or not. For the sake of simplicity we continue to focus on the $\mathcal{E} = 0$ marginally bound case.

The general solution for $r(R, t)$ when $\mathcal{E} = 0$ is \cite{Giesel:2023hys}
\beq \label{solR}
r(R,t) = \left( 2Gm \left( \f{9}{4} (t - \alpha)^2 + \Delta \right) \right)^{1/3},
\ee
where $m(R)$ is given by the initial conditions, as is $\alpha(R)$ which is a constant of integration.

Imposing the initial condition that $r(R, t_0) = R$, the solution \eqref{solR} can be inverted to solve for $\alpha$,
\begin{equation} \label{alpha}
\alpha(R)= t_0 + \sqrt{\f{4}{9} \left( \frac{R^{3}}{2 G m(R)}-\Delta \right)},
\end{equation}
where the positive sign in front of the square root is selected for dust that is initially collapsing (rather than expanding). Note that by a suitable rescaling it is always possible, as done here, to choose the radial coordinate $R$ to initially agree with the areal radius $r$ at $t=t_0$ (assuming that the areal radius is initially monotonically increasing; this will always be the case for the configurations we consider)---this choice will simplify the analysis.  Of course, the relation $r(R, t_0) = R$ only holds at the initial time $t_0$.

(As an aside, note that it may be more appropriate to give initial conditions in terms of $m(r)$, not $m(R)$, since $r$ has a clear geometric meaning while $R$ is a radial coordinate that can be freely rescaled; however, since we are choosing $R$ such that it initially agrees with $r$, we will give initial conditions in terms of $R$ instead, using this identification.)

The two conditions necessary for a shell-crossing singularity to occur are $\partial_R m \neq 0$ and $\partial_R r = 0$. The first can be checked directly from the initial profile $m(R)$, while a direct calculation from \eqref{solR} shows that $\partial_R r = 0$ is equivalent to the condition
\beq \label{eqt}
(t-\alpha)^2 \partial_{R}m - 2m (t-\alpha) \partial_{R}\alpha + \f{4\Delta}{9} \partial_{R}m=0.
\ee
For any $R$ such that $\partial_R m \neq 0$, if there is any $t_s$ that satisfies this equality, then there will necessarily be a shell-crossing singularity at the spacetime point $(R, t_s)$.

To see if there is a real solution for $t$, solving the quadratic equation gives
\begin{equation} \label{solt}
t - \alpha = \frac{m\partial_{R}\alpha}{\partial_{R}m} \left( 1 \pm \sqrt{1-\frac{4\Delta (\partial_{R}m)^{2}}{9 m^{2}(\partial_{R}\alpha)^{2}}} \right),
\end{equation}
therefore, there exists at least one real solution for $t$ if and only if
\beq \label{cond}
\f{m | \partial_R \alpha|}{\partial_R m} \ge \f{2 \sqrt\Delta}{3}.
\ee
If this condition (which is independent of $t$), together with $\partial_R m \neq 0$ (also independent of $t$), is satisfied for any $R$, then a shell-crossing singularity will necessarily occur.

Note that if there are real solutions to \eqref{solt} for $t-\alpha$, they must both have the same sign, which is determined by $\partial_R \alpha$ since $m$ and $\partial_R m$ are both positive (assuming positive energy density everywhere).  At the radial coordinate $R$, a bounce locally occurs at the time $t = \alpha$, so a positive solution for $t-\alpha$ (i.e., with $\partial_R \alpha > 0$) corresponds to a shell-crossing singularity that occurs after the bounce, while negative solution for $t - \alpha$ (i.e., with $\partial_R \alpha < 0$) indicates that a shell-crossing singularities forms before the bounce.

Finally, a constraint on the time when a shell-crossing singularity occurs is given by rewriting \eqref{eqt} as
\beq
\label{30}
\f{2(t-\alpha)}{(t-\alpha)^2 + \tf{4\Delta}{9}} = \f{\partial_R m}{m \partial_R \alpha}.
\ee
It is straightforward to verify that the extrema of the function $2(t-\alpha) / [(t-\alpha)^2 + 4\Delta/9]$ are $-3/(2 \sqrt\Delta)$ and $3/(2 \sqrt\Delta)$, and further that the extrema, and every point between, is attained by that function for $t-\alpha \in [-2 \sqrt\Delta / 3, 2 \sqrt\Delta / 3]$, i.e., of the order of a Planck time either side of the (local) bounce at $t = \alpha$. As a consequence, if a shell-crossing singularity does form it will occur at the latest at
\beq \label{tmax}
t_{\rm latest} = \alpha + \f{2 \sqrt\Delta}{3},
\ee
within a time $2\sqrt\Delta/3$ after the bounce.

\subsection{Two simple models}

We will now consider two simple models that are both initially arbitrarily close to the OS profile, and show that shell-crossing singularities occur in both.

\subsubsection{First model}

The first model we consider is based on defining $\rho(t_0)$ in three pieces: an innermost region $R<R_1$ where the energy density is a constant $\rho_o$, an intermediate region where $\rho$ decreases linearly to 0, and an outer region where $\rho$ vanishes, namely 
\beq \label{model1}
\rho = 
\begin{cases}
\displaystyle \rho_o, & {\rm for~} R < R_1, \\[3pt]
\displaystyle \rho_o \cdot \f{ R_2 - R }{R_2 - R_1}, \quad & {\rm for~} R_1 < R < R_2, \\[3pt]
0, & {\rm for~} R > R_2.
\end{cases}
\ee
By construction, $\rho(t_0)$ is everywhere continuous (although not differentiable) in this model. Further, this initial configuration can be made arbitrarily close to the OS case, for $R_2-R_1$ sufficiently small. For the sake of concreteness, consider the $L^1$ norm of $\rho(t_0) - \rho_{OS}(t_0)$, choosing the parameters of the OS initial profile to have the same $\rho_o$, and taking $R_{OS} = R_1$. Then,
\begin{align}
\| \rho(t_0) - \rho_{OS}(t_0) \|_1 &= \int_0^\infty \!\! \dd R ~ |\rho(t_0) - \rho_{OS}(t_0)| \nn \\
&= \f{1}{2} \rho_o (R_2 - R_1),
\end{align}
which can clearly be made arbitrarily small by taking $R_2 - R_1$ to be as close to 0 as necessary.  (Note that one could choose instead to define the $L^1$ norm using the integral over $\mathbb{R}^3$, but in any case the resulting norm of $\rho(t_0) - \rho_{OS}(t_0)$ can be made arbitrarily small by taking $R_2 - R_1$ arbitrarily close to zero.)

In Sec.~\ref{s.os-ch}, we saw that no shell-crossing singularities will form in the homogeneous inner region (and $\partial_R m = 0$ in the outer region, so there can be no shell-crossing singularities there either), so we will focus on the intermediate region $R_{1}<R<R_{2}$.

In the intermediate $R_1 < R < R_2$ region, the mass function $m$ is
\begin{equation}
m=\frac{4 \pi \rho_o }{3(R_{2}-R_{1})} \left( R_2 R^3 -\f{R_{1}^{4}}{4} -\f{3R^{4}}{4} \right).
\label{mass1}
\end{equation}
Since $\partial_{R}m>0$ for $R \in(R_{1},R_{2})$, if $\partial_{R}r(t,R)=0$ for some $R$ in that range, and for any $t$, a shell-crossing singularity occurs at that point. A direct calculation shows that the condition \eqref{cond} for the occurrence of a shell-crossing singularity at $R$ for this model gives
\beq
\! \f{1}{\sqrt{\f{R^3}{2Gm} - \Delta}}
\left| \f{R_2 - R_1}{R_2 - R} - \f{R_2 - R_1}{R_2 - \tf{3}{4} R - \tf{R_1^4}{4R^3}} \right|
\ge \sqrt \f{\rho_o}{\rho_c},
\ee
where $\rho_c = 3 / (8 \pi G \Delta)$ is the critical energy density of loop quantum cosmology.

Note that no matter the values of $R_1, R_2$ and $\rho_o$, this inequality will be satisfied for $R$ (smaller than but) sufficiently close to $R_2$ for the denominator of the first fraction inside the absolute values to become sufficiently large for the inequality to hold.

As a result, a shell-crossing singularity will necessarily form, shortly after the bounce (since $\partial_R \alpha > 0$) for any set of initial data of the form \eqref{model1}, including initial choices for $\rho(R, t_0)$ arbitrarily close to the OS configuration.

\subsubsection{Second model}

The second model we consider is a further modification of the OS profile.  We again define the initial $\rho(t_0)$ in three pieces: an innermost region where $\rho$ decreases linearly from a maximal value $\rho_m$ at the origin to a smaller value $\rho_1$ at $R = R_1$, an intermediate region where $\rho$ decreases linearly, although at a different rate, to zero at $R=R_2$, and vanishes in the outer region,
\beq
\rho =
\begin{cases}
\displaystyle \rho_m \cdot \f{ R_1 - a R }{R_1}, & {\rm for~} R < R_1, \\[8pt]
\displaystyle \rho_1 \cdot \f{ R_2 - R }{R_2 - R_1}, \quad & {\rm for~} R_1 < R < R_2, \\[8pt]
0, & {\rm for~} R > R_2,
\end{cases}
\ee
where $0 < a < 1$ and $\rho_1 = (1-a) \rho_m$.  This initial configuration can also be made arbitrarily close to an OS model (with parameters $\rho_o = \rho_m$ and $R_{OS} = R_1$), since $\| \rho(t_0) - \rho_{OS}(t_0) \|_1 = [ aR_1 + \rho_1 (R_2- R_1)] /2$ can be made arbitrarily small by taking $a$ and $R_2-R_1$ arbitrarily close to 0.

We will focus on the innermost region $0< R < R_1$; the mass function $m$ in that region is
\begin{equation}
m=\frac{4 \pi \rho_{m} R^{3}}{3} \left( 1 - \f{3aR}{4R_{1}} \right),
\label{mass2}
\end{equation}
and the solution for $\alpha$ is
\begin{equation}
\alpha(R)= t_0 + \sqrt{\frac{2R_{1}}{3 \pi G \rho_{m}(4R_{1}-3aR)}-\frac{4\Delta}{9}},
\end{equation}
as usual choosing the positive root for a contracting initial profile.

Since $\partial_R m \neq 0$ in the innermost region, the remaining condition for a shell-crossing singularity to form is that the inequality \eqref{cond} be satisfied, which for this model is equivalent to
\beq
\f{aR}{8 R_1} \cdot \f{\left[(1 - \f{aR}{R_1})\sqrt{1 - \f{3aR}{4R_1}}\right]^{-1}}{\sqrt{ 1 - \f{\rho_m}{\rho_c} (1 - \f{3aR}{4R_1})}} \geq \sqrt\f{ \rho_m}{\rho_c},
\ee
where $\rho_c = 3 / (8 \pi G \Delta)$ is the critical energy density of loop quantum cosmology; note that the left side of the inequality is maximized at $R = R_1$.

In this case, if $a$ is sufficiently small, there will not be any shell-crossing singularities in the innermost region (and in particular, for the limiting case of a homogeneous profile with $a=0$). Note however that to avoid shell-crossing singularities, $a$ must be very small.  Assuming that the initial profile for $\rho(t_0)$ is such that $\rho_m \ll \rho_c$, $a \ll 1$, the condition that there not be any shell-crossings at radius $R$ is $aR/R_1 < 8 \sqrt{\rho_m / \rho_c}$. Clearly, for initial profiles with $\rho_m \ll \rho_c$, this is a strong constraint on $a$.

For this second model, shell-crossing singularities can occur close to the origin in the innermost region, with larger value of $a$ ensuring that shell-crossing singularities will occur closer to the origin.  Also, even if $a$ is chosen to be sufficiently small so that a shell-crossing singularity does not occur in the innermost region $R<R_1$, shell-crossing singularities will necessarily occur in the intermediate region $R_1 < R < R_2$, whose initial energy density profile is identical to the intermediate region of the first model, and where it was shown that a shell-crossing singularity will necessarily form.

\subsection{General initial profiles with $\mathcal{E}=0$}

There is a very large class of profiles that will eventually lead to the formation of a shell-crossing singularity. As shall be shown here, for the case $\mathcal{E} = 0$, the condition for a shell-crossing singularity to form is that the dust energy density be sufficiently inhomogeneous, or go sufficiently close to zero. Importantly, this includes all initial (continuous) profiles for $\rho(R, t_0)$ that are of compact support (and with non-vanishing $m$, i.e., non-Minkowski): every such initial profile will inevitably lead to the formation of a shell-crossing singularity.

To determine whether a shell-crossing singularity will occur at some radial coordinate $R=R_1$, it is convenient to introduce the average initial energy density from the origin to $R_1$, defined as
\beq
\bar \varrho_1 = \f{3 m(R_1)}{4 \pi R_1^3},
\ee
and to express the initial energy density at $R_1$, given by $\varrho(R_1) = \rho(R_1, t_0) = \partial_R m(R_1) / (4 \pi R_1^2)$, as
\beq
\varrho_1 := \varrho(R_1) = \bar \varrho_1 + \delta \varrho_1.
\ee
Note that we use the notation $\varrho$ for these quantities since they depend only on $R$, unlike the energy density $\rho$ which of course is dynamical.  Note also that all of these quantities depend on the radial coordinate $R_1$; in particular, $\bar \varrho_1$ is the average initial energy density from the origin to $R_1$, so it also depends on $R_1$.

From these definitions, it follows that
\beq
\f{\partial_R m}{m} \Bigg|_{R = R_1} = \f{3 \varrho_1}{R_1 \bar \varrho_{1}},
\ee
and (for $\mathcal{E} = 0$) $\partial_R \alpha$ can be calculated directly from \eqref{alpha},
\beq
\partial_R \alpha = \f{R^2}{2 G m \sqrt{ \f{R^3}{2 G m} - \Delta } } \left( 1 - \f{R \partial_R m}{3 m} \right),
\ee
which, for $R=R_1$, can be rewritten as
\beq
\partial_R \alpha(R_1) = - \f{ \sqrt\Delta}{R_{1}} \cdot \f{\delta \varrho_1}{\bar \varrho_1} \sqrt \f{\rho_c}{\bar \varrho_1} \f{1}{\sqrt{1 - \f{\bar \varrho_1}{\rho_c}}}.
\ee

Combining these calculations, at $R=R_1$
\beq
\f{m \partial_R \alpha}{\partial_R m} = - \f{\sqrt\Delta}{3} \f{\delta \varrho_1}{\varrho_1} \left[ \sqrt{ \f{\bar \varrho_1}{\rho_c} \left(1 - \f{\bar \varrho_1}{\rho_c} \right) } \,\, \right]^{-1},
\label{condition}
\ee
where $\rho_c = 3 / (8 \pi G \Delta)$ is the critical energy density of loop quantum cosmology.

As a result, the condition \eqref{cond} becomes
\beq \label{ineq}
\f{| \delta \varrho_1 |}{\varrho_1} \ge 2 \sqrt{ \f{\bar \varrho_1}{\rho_c} \left(1 - \f{\bar \varrho_1}{\rho_c} \right) }.
\ee
For any radial coordinate $R$ where $\partial_R m \neq 0$, if either $|\delta \varrho_1|$ is sufficiently large or $\varrho_1$ is sufficiently small, so that \eqref{ineq} is satisfied, then a shell-crossing singularity will necessarily form, at the latest at $t = \alpha + 2 \sqrt\Delta / 3$, as shown in \eqref{tmax}.

We emphasize that it is easy to find initial conditions such that this condition is satisfied; in particular, it is satisfied for any initial configuration with non-zero $m$ where the initial energy density is continuous and of compact support.

\begin{theorem*}
For the case $\mathcal{E} = 0$, a shell-crossing singularity forms if the initial distribution of the dust energy density $\rho(R, t_0)$ is non-negative, continuous, of compact support, and for which $m(R)$ is not everywhere zero.
\end{theorem*}

\noindent
{\it Proof:}
Since $m(R)$ is not everywhere zero, there must exist some finite open interval $(R_1, R_2)$ where the non-negative function $\rho(R, t_0)$ is strictly positive and $\partial_R m > 0$. Further, since $\rho(R, t_0)$ has compact support, by definition it is possible to find such an interval where $R_2$ satisfies the condition that $\rho(R_2, t_0) = 0$. Then, by continuity it is possible to find a positive and arbitrarily small $\rho(R, t_0)$ for some $R \in (R_1, R_2)$ (in particular by taking $R$ arbitrarily close to $R_2$), so it also follows that $\varrho(R)$ can also be made arbitrarily small. Since $\varrho(R)$ can be made arbitrarily small with $R<R_2$ and $m$ is a non-decreasing function, it is possible to ensure that $\varrho(R) < 3 m(R) / (4 \pi R^3) = \bar \varrho(R)$, so that $\delta \varrho(R)$ is negative, and $|\delta \varrho(R)|$ is bounded below as $R$ is chosen to make $\varrho(R)$ arbitrarily small. Therefore, the denominator of $|\delta \varrho(R)| / \varrho(R)$ can be made arbitrarily small while the numerator is bounded below, guaranteeing that the inequality \eqref{ineq} is satisfied for some $R \in (R_1, R_2)$. It follows that $\partial_R r = 0$ will occur at some time for that $R$, and since $\partial_R m \neq 0$ for all $R \in (R_1, R_2)$, therefore when $\partial_R r = 0$ a shell-crossing singularity will occur. $\Box$

\bigskip

Of course, the constraint \eqref{ineq} holds much more generally than just for profiles of compact support---it can also be applied to non-compact configurations, although in this case it is necessary to check whether the inequality holds on a case by case basis for each choice of initial conditions. Given that the right side is suppressed by a factor of $1/\sqrt{\rho_c}$, it is clear that it is not difficult to find initial conditions for which a shell-crossing singularity will occur, whether the initial profile for the energy density of the dust field is compact or not.

\subsection{The dust-time gauge}

As we have seen, relaxing the areal gauge is not sufficient to avoid shell-crossing singularities when considering models beyond Oppenheimer-Snyder---in fact, at all times up until the shell-crossing singularity occurs, it can be verified that the areal gauge holds (and the solutions to the equations of motion in comoving coordinates cannot be used beyond the shell-crossing singularity). But what about relaxing the dust-time gauge?

One may hope that by using a different time coordinate, it may be possible to avoid characteristics crossing. This may be the case in vacuum as argued in \cite{Fazzini:2023scu}, where caustics in the coordinate system could just indicate the failure of the coordinates, and in such a situation it is possible to simply use a different set of coordinates. On the other hand, shell-crossing singularities (although weak) are curvature singularities, with curvature scalars diverging---and curvature singularities cannot be avoided by introducing a new coordinate system.

\section{Physics of shocks in gravitational collapse}
\label{s.shock}

The perspective from fluid dynamics, where shocks are a common and well-understood phenomenon provides some insight for shock formation in gravitational collapse. Indeed,  it has recently been suggested that shocks may occur in  analog gravity models where the medium's energy density becomes sufficiently concentrated, and this too may yield insights for shock formation in quantum gravity \cite{Braunstein:2023jpo}.

Fluid dynamics is a macroscopic continuum approximation of the collective motion of a large number of molecules described by coarse-grained quantities such as density $\rho$ and pressure $p$ that are normally treated as continuous functions of space and time. Coarse-graining presupposes the existence of regions that contain sufficient numbers of molecules, and it is possible that adjacent regions give significantly different values for these quantities. Thus the assumption of continuity of $\rho$ and $p$ is not always justified. Shocks are emergent macroscopic discontinuities, and do not reflect any underlying microscopic discontinuity. 

With this perspective, shocks are not surprising in quantum gravity if one assumes that spacetime is emergent from a large number of geometry quanta, as proposed for instance in LQG.  Conversely, the appearance of shocks in classical or effective LQG gravity (as discussed above) provides an argument for a hydrodynamic picture of spacetime with a discrete underlying microstructure. 

In general, a weak solution is one where one or more coarse-grained observables are not differentiable, as summarized in the Introduction. Although a partial differential equation has a unique differentiable solution on a specified domain (if it satisfies some appropriate conditions), this is not the case for integral equations, where the solution need not be continuous or differentiable on the domain. Weak solutions are not unique because they depend on the choice of the field in terms of which the integral form of the equations of motion are expressed \cite{PDEbook}, in particular non-linear redefinitions of the field lead to shocks propagating at different speeds. Hence it is important to select appropriate ``fundamental" fields for the integral equations. This requires additional intuition from the underlying physics (quantum gravity in this case).
 
While this lack of uniqueness for weak solutions may initially seem disconcerting, the choice of  fundamental fields contains information about the underlying physical constituents of the coarse-grained observable: in  fluid dynamics the physical ingredient may be the conservation of mass or particle number, which indicates that the appropriate field for expressing the integral form of dynamics is mass or number density \cite{PDEbook}. As a result, in weak solutions, microscopic degrees of freedom can have a significant impact on the macroscopic dynamics---this is of particular interest for quantum gravity where guidance from empirical observation is very limited.

For LTB models of gravitational collapse, this raises two questions. The first concerns the physics underlying the assumption of vanishing pressure: should zero pressure be assumed in all situations, or as an approximation that holds unless particle shells approach each other?  In the first case, it may be natural to select weak solutions where shells pass directly through each other without resistance if a shell crossing occurs (see, e.g., the analysis in \cite{Dray:1985yt});  in the second case, which seems more realistic, it is natural to expect resistance to shells passing through each other (due to pressure not being negligible at sufficiently high particle density), and thus leading to shells piling up and forming a shock.

The second question for selecting the physically correct weak solution concerns the appropriate dynamics for shocks: what is the microscopic input from quantum gravity that determines the physically appropriate weak solution for gravitational collapse? From the perspective of LQG, it seems natural to expect that (the integral form) of the dynamics use the Ashtekar-Barbero variables (as in \cite{Husain:2021ojz, Husain:2022gwp}).  However, other choices may be of interest. This is a potentially promising question to explore since it may provide more insight into the connection between microscopic degrees of freedom in LQG and the emergent macroscopic dynamics of shocks.

A final point to consider is whether the effective solutions remain valid up to the point where characteristics cross, and whether the weak effective solutions discussed in \cite{Husain:2021ojz, Husain:2022gwp} are valid beyond such a shell-crossing. While a detailed answer to this question requires further work, the comparison to fluid mechanics is illuminating. In fluid mechanics, at characteristic crossings it does not become necessary to consider the atomistic or quantum nature of the fluid's constituents to determine the dynamics (at least if we are only probing the system at length scales much larger than the atomic scale); rather, it is sufficient to find weak solutions to the dynamics. This comparison suggests that the same may reasonably be expected to be true here: at length scales much larger than $\lp$, weak solutions to the integrated effective equations can be used, and a recourse to fully quantum equations does not seem necessary.

\newpage

\section{Discussion}
\label{s.disc}

Shell-crossing singularities occur in the  LQG effective dynamics for marginally bound ($\mathcal{E} = 0$) LTB spacetimes for initial distributions of the energy density $\rho(t_0) \ge 0$ that are either (i) sufficiently inhomogeneous or (ii) become sufficiently close to zero at a point, including all continuous profiles of compact support.

These shell-crossing singularities indicate the failure of comoving coordinates due to characteristics crossing, and therefore show the necessity of seeking weak solutions. In turn, the weak solutions that arise beyond the point of characteristic crossing are shock waves. 

Our results stand in contrast to recent claims \cite{Fazzini:2023scu, Giesel:2023hys} that shocks in dust collapse in effective LQG are coordinate artefacts; these claims are based on analyses focused on the OS model, and the implicit assumption that it is a representative example. We have shown that this is not the case by exhibiting families of initial energy density profiles (see models 1 and 2 in Sec.~\ref{s.beyond}) that are  arbitrarily close to OS data for which shell-crossing singularities necessarily occur.

Since comoving coordinates break down at shell-crossing singularities, it becomes necessary to use other coordinates past such crossings. The areal gauge still holds at the time the first shell-crossing singularity forms---this is because the singularity initially forms precisely at the first point $R_s$ where $\partial_R r = 0$ (and this derivative remains strictly positive either side of that point, since by definition this is the first point where $\partial_R r$ vanishes), so $R_s$ is an inflexion point for $r$ (not an extremum), and $r$ also increases everywhere else. In particular, since $r$ remains an increasing function at the shell-crossing singularity, the areal gauge can be used at such points. This suggests that it is possible to use the integral form of the non-linear wave equation in the areal gauge to find the appropriate weak solutions, as proposed in \cite{Husain:2021ojz, Husain:2022gwp}. Nevertheless it would be valuable to use the general framework developed in \cite{Giesel:2023tsj} to find other coordinates (that are not comoving and do not use the areal gauge) to study the weak solutions that are expected to hold beyond the shell-crossing singularity. 

In summary, we have shown that characteristic crossing and the formation of shell-crossing singularities commonly occur for the PDEs that describes LTB dust collapse in effective LQG; the inevitable consequence of this fact is the necessity of seeking weak solutions.

\acknowledgments

\noindent
This work was supported in part by the Natural Sciences and Engineering Research Council of Canada.

\newpage

\appendix

\section{Generalized Painlev\'e-Gullstrand coordinates in LTB spacetimes}
\label{app.pg}

The effective LQG dynamics for the LTB spacetime can be derived in terms of the generalized Painlev\'e-Gullstrand coordinates by imposing the areal and dust-time gauge, and then performing a loop quantization on the resulting gauge-fixed system, for details see \cite{Kelly:2020lec}.

The metric then has the form
\beq
\dd s^2 = - \dd t^2 + \f{(E^b)^2}{r^2} \Big( \dd r + N^r \dd t \Big)^2 + r^2 \dd\Omega,
\ee
where the shift vector is
\beq \label{app-N}
N^r = - \f{r}{\sqrt\Delta} \sin \f{\sqrt\Delta b}{r} \cos \f{\sqrt\Delta b}{r},
\ee
and $b$ and $E^b$ are, respectively, the remaining components of the Ashtekar-Barbero connection and of the densitized triad, and are canonically conjugate
\beq
\{b(r_1, t), E^b(r_2, t)\} = G \, \delta(r_1 - r_2).
\ee

The dynamics are generated by the physical Hamiltonian (which is not a constraint after the gauge-fixing)
\begin{align}
\mathcal{H} =& \, \f{1}{2G} \Bigg[ \f{E^b}{r} \partial_r \left( \f{r^3}{\Delta} \sin^2 \f{\sqrt\Delta b}{r} \right) - \f{3r}{E^b} \nn \\ & \qquad\qquad 
+ \f{2r^2}{(E^b)^2} \partial_r(E^b) + \f{E^b}{r} \Bigg],
\end{align}
giving the equations of motion
\begin{gather}
\dot b = - \f{1}{2 \Delta r} \partial_r \left( r^3 \sin^2 \f{\sqrt\Delta b}{r} \right) + \f{1}{2} \left( \f{r}{(E^b)^2} - \f{1}{r} \right), \\
\dot E^b = - \f{r^2}{\sqrt\Delta} \sin \f{\sqrt\Delta b}{r} \cos \f{\sqrt\Delta b}{r} \, \partial_r \left( \f{E^b}{r} \right),
\end{gather}
and the energy density of the dust field is
\beq
\rho = - \f{\mathcal{H}}{4 \pi r E^b}.
\ee

These equations can be simplified by the redefinition of $E^b$ in terms of $\mathcal{E}$ through
\beq
E^b = \f{r}{\sqrt{1 + \mathcal{E}}},
\ee
and then the equations of motion become
\begin{gather}
\label{app-b}
\dot b = - \f{1}{2 \Delta r} \partial_r \left( r^3 \sin^2 \f{\sqrt\Delta b}{r} \right) + \f{\mathcal{E}}{2r}, \\
\label{app-e}
\dot{\mathcal{E}} = - \f{r}{\sqrt\Delta} \sin \f{\sqrt\Delta b}{r} \cos \f{\sqrt\Delta b}{r} \, \partial_r \mathcal{E},
\end{gather}
while in terms of $\mathcal{E}$ the energy density is
\beq
\rho = \f{1}{8 \pi G r^2} \partial_r \left( \f{r^3}{\Delta} \sin^2 \f{\sqrt\Delta b}{r} - r \mathcal{E} \right),
\ee
motivating the definition of the gravitational mass
\begin{align} \label{app-def-m}
m(r) & \, = 4 \pi \int_0^r \dd \tilde r ~ \tilde r^2 \rho(\tilde r) \nn \\
& = \f{1}{2G} \left( \f{r^3}{\Delta} \sin^2 \f{\sqrt\Delta b}{r} - r \mathcal{E} \right).
\end{align}

Using the equations of motion for $b$ and $\mathcal{E}$,
\beq \label{app-m}
\dot m = - \f{r}{\sqrt\Delta} \sin \f{\sqrt\Delta b}{r} \cos \f{\sqrt\Delta b}{r} \, \partial_r m.
\ee
The PDEs \eqref{app-e} and \eqref{app-m} are precisely those given in \eqref{m-pde}-\eqref{e-pde}, with $f = -N^r$, see \eqref{app-N}.

To recover the characteristic equations for the LTB spacetime, it is sufficient to introduce characteristic curves $(r(s), t(s))$, whose parametric equations satisfy
\beq
\f{\dd t}{\dd s} = 1, \qquad \f{\dd r}{\dd s} = \f{r}{\sqrt\Delta} \sin \f{\sqrt\Delta b}{r} \cos \f{\sqrt\Delta b}{r},
\ee
so the dynamics for $m$ and $\mathcal{E}$ along the characteristics trivialize to
\beq
\f{\dd m}{\dd s} = 0, \qquad \f{\dd \mathcal{E}}{\dd s} = 0.
\ee
Using \eqref{app-def-m} to express the characteristic equation for $r(t)$ in terms of $m$ and $\mathcal{E}$ gives
\beq
\dot r^2
= r^2 \left( \f{2Gm}{r^3} + \f{\mathcal{E}}{r^2} \right) \left[ 1 - \Delta \left( \f{2Gm}{r^3} + \f{\mathcal{E}}{r^2} \right) \right].
\ee
These are precisely Eqs.~\eqref{e-m-dot}--\eqref{rdot} derived using the comoving coordinate $R$. Note also that the equation for the gravitational mass $m$ expressed in terms of $R$ \eqref{def-m} follows from changing variables from $r$ to $R$ in \eqref{app-def-m}. As emphasized in the text, these equations derived using the method of characteristics only hold so long as characteristic curves do not cross.

Finally, as mentioned at the end of Sec.~\ref{s.ltb}, despite the relative simplicity of the equations of motion for $m$ and $\mathcal{E}$ (and their similarity to each other), these are not the ideal choice of fundamental variables since together they do not determine the sign of the velocity of the fields given by $-N^r$, rather the variable $b$ is needed for this. Because of this, even though the equation of motion for $b$ is slightly more complicated, it is nonetheless necessary to look for weak solutions (allowing for the possibility of characteristics crossing) for $b$ and $\mathcal{E}$, and once these quantities are known then it is straightforward to calculate $m$ from \eqref{app-def-m}; this has been done both for the case $\mathcal{E} = 0$ \cite{Husain:2021ojz, Husain:2022gwp} and $\mathcal{E} \neq 0$ \cite{Cipriani:2024nhx}.

\newpage

\raggedright

\end{document}